\documentclass[runningheads]{llncs}
\usepackage{graphicx}
\usepackage{subfigure}

\begin{document}
\title{Non-Binary Gender Expression in Online Interactions}
%
%
\author{Rebecca Dorn\and 
Negar Mokhberian\and
Julie Jiang\and
Jeremy Abramson\and
Fred Morstatter\and
Kristina Lerman}
\authorrunning{R. Dorn et al.}
%
\institute{University of Southern California, Information Science Institute\\
\email{rdorn,nmokhber,juliej,abramson,fredmors,lerman@isi.edu}}
\maketitle              
\begin{abstract}

Many openly non-binary gender individuals participate in social networks. However, the relationship between gender and online interactions is not well understood, which may result in disparate treatment by large language models. We investigate individual identity on Twitter, focusing on gender expression as represented by users’ chosen pronouns . We find that non-binary groups tend to receive less attention in the form of likes and followers. 
We also find that non-binary users send and receive tweets with above-average toxicity. The study highlights the importance of considering gender as a spectrum, rather than a binary, in understanding online interactions and expression.

\end{abstract}
\section{Introduction}
An individual’s identity, defined along multiple dimensions such as age, race, ethnic origin, socio-economic status, etc., influences self expression and social interactions ~\cite{deaux1993reconstructing}. 
As social life has migrated online, those influences increasingly play out in social media.
%
Our paper focuses on gender, an integral dimension of individual identity. The concept of gender has changed in recent years, moving from the traditional binary  `male'/`female' classification to an understanding that gender forms a spectrum. Gender minorities, including LGTBQ+ people, are at a higher risk of harassment, workplace discrimination, social isolation and shortened lifespan~\cite{tabaac2018discrimination}. LGTBQ+ youth 
are increasingly active in online communities~\cite{mcinroy2019lgbtq}. However, how their posts are received by others has not been systematically examined.  

As a proxy of gender identity, we study pronouns users choose to display in their online profile or biography. 
These pronouns range from the traditional binary gender categories, such as `she/her' 
and `he/him',
to non-binary and gender nonconforming~\cite{hicks-etal-2016-analysis} categories `they/them', `she/ze', `she/they/xe', etc. 
We study the eight pronoun groups that contain the most members, revealing the spectrum of gender identity online (See §3 for more details).

We study how gender identity mediates online expression and interactions, using state-of-the-art tools to assess toxicity of language used in posts and replies. 
We investigate the following research questions:
%


\begin{description}
    \item[RQ1] How do users in different pronouns groups vary in their level of online activity and the online attention they 
    receive from others?  
    receive more negative replies? 
    \item[RQ2] Which user pronoun groups convey more toxicity on Twitter? Which groups experience more toxicity from others?
\end{description}

We find that while binary and non-binary groups are equally active online, they differ in the amount of attention they receive
and their messages’ toxicity and use of profanity. We hope that our illustration of the need to treat gender expression as a spectrum and not a binary category contributes to the improvement of computational methods that analyze gender equity and equality.
To make the online space more inclusive and welcoming to all people, it is important to understand potential interactions between gender identity and online expression.

\section{Related Works}

\paragraph{Gender and Identity}
Identity is a construct used by sociologists and social psychologists to help understand how people define themselves and experience others in social interactions (e.g., \cite{deaux1993reconstructing}). Gender is a core dimension of identity. Inspired by second-wave feminism \cite{article}, people have started to draw 
a distinction between sex (biologically-produced) and gender (culturally-produced) identity. 
This helped resolve the tension between the traditional conceptualization of gender in Western society and science as binary (i.e., `male' and`female'), and historical and societal evidence of the presence of non-binary individuals \cite{genderqueer}. 
Despite the growing awareness of this distinction, computational social science and machine learning typically treat gender as binary. For example, machine learning systems in recent years have claimed to predict an individual as male or female based on their written name, handwriting, voice and other characteristics \cite{siddiqi2015automatic,raahul2017voice,hu2021s}. As another example, Wordnet was shown to under-represent non-binary gender terms \cite{hicks-etal-2016-analysis}. Despite offering non-binary gender options, Meta has been found to internally reconfigure user genders according to their perceived assigned gender at birth \cite{doi:10.1177/1461444815621527}. This results in continuing a legacy of erasure and harm for gender-queer people \cite{10.1145/3274357}. 

Twitter biographies have been used to measure expressions of personal identity and cultural trends. Previous work introduced Longitudinal Online Profile Sampling (LOPS) which measures identity formation through the evolution of a user's Twitter biography \cite{9355032}. LOPS was used to compare how 1.35 million Twitter user biographies evolved over 5 years, taking one snapshot of user biographies annually \cite{jones2021dataset}. The longitudinal study found that the tokens with the highest prevalence within biographies were \textit{he}, \textit{him}, \textit{she} and \textit{her}. The LOPS studies relied upon the notion of \textbf{personally expressed identity} (PEI) where individuals declare \textit{their own} attributes.





\paragraph{Toxicity Detection}
\label{sec:toxicity}
Harmful speech, a category that includes personal attacks, insults, threats, harassment, and hate speech targeting specific subgroups, contributes to the toxicity of online communities~\cite{massanari2017gamergate}. 
To combat negative effects of harmful speech, researchers developed methods for toxicity detection. 
One popular method is the Perspective API~\cite{perspective}, which has been trained on diverse social media posts that human annotators labeled along several dimensions of harm, such as identity attack, insult, obscenity, and threat. 
However, Perspective API was shown to be biased, giving texts referring to certain racial or gender groups higher toxicity scores \cite{hutchinson2020social}. 
To combat bias, \cite{hanu2020unitary} introduced Detoxify API, which was trained on open source data emphasizing toxicity towards specific identities ~\cite{conneau2019unsupervised}.
We use the latter state-of-the-art method to measure toxicity of speech made by members of different gender identity groups, as well as speech directed at them.



\section{Methods}
\paragraph{Data}
\label{data}
Our work is supported by a collection of over 2 billion tweets related to the Covid-19 pandemic collected between January 21, 2020, and November 5, 2021 \cite{Chen_2020}. As most platform-engaged Twitter users tweeted about Covid-19 at some point, this generates a sample of Twitter users who are active.
This data includes tweets from 2,066,165 users with specified pronouns in their Twitter profiles or biographies. The presence of pronouns is determined by whether a user has specified any combination of $\{$he, him, his, she, her, hers, they, them, theirs, their, xe, xem, ze, zem$\}$ separated by forward slashes or commas, with any or no white space in their profile descriptions~\cite{jiang2022your}. Our data was collected in real-time continuously, meaning we record the profile descriptions of the users at the time of their first tweets.


\paragraph{Gender Spectrum}
Gender identity (e.g., woman) is separate from the pronouns a person uses (e.g., she/hers or they/them). We use the term gender expression in reference to a person’s identity surrounding gender, not the gender label assigned. We use the term
\textbf{non-binary} to describe anyone who does not identify as either a man or a woman. To quantify differences by gender we are required to operationalize gender. We use different user pronouns to classify users according to their gender expression, recognizing that these are proxy measures that do not definitely establish whether someone is a man, a non-binary person, etc.  

We group the pronouns into five different series: she/her/hers, he/him/his, they/them/theirs, xe/xem and ze/zem. We encode the different combinations of pronouns used via a 5-digit dummy variable that is malleable to a range of gender representations and computationally efficient. 

We encode the pronouns of all ~2 million users into this 5-digit schema. We then identify all pronoun groups with at least 1,000 members. We randomly sample up to 600 users from each group with a public Twitter profile as of September 30, 2022, excluding users who eliminated their public profiles before our analysis was complete. This gives us 8 pronoun groups with at least 350 valid users, as shown in Table \ref{tab:sample}, which lists groups in decreasing order of their size within the COVID-19 data set.

For each user in our sample, we collect at most 1,000 of their most recent tweets posted before 
September 30, 2022. 
Tweets are retrieved using the API's \verb|user timeline|
call. 
Table \ref{tab:sample} reports total tweets in sample authored by each pronoun group.

\begin{table}[]
\centering
\footnotesize
\begin{tabular}{cccc}
\hline
\multicolumn{1}{|c|}{\textbf{Group}} & \multicolumn{1}{c|}{\textbf{Original Users}} & \multicolumn{1}{c|}{\textbf{Sample Users}} & \multicolumn{1}{c|}{\textbf{Sample Tweets}} \\ \hline
\multicolumn{1}{|c|}{She}            & \multicolumn{1}{c|}{1,194,565}               & \multicolumn{1}{c|}{508}                   & \multicolumn{1}{c|}{464,262}                \\ \hline
\multicolumn{1}{|c|}{He}             & \multicolumn{1}{c|}{461,264}                 & \multicolumn{1}{c|}{559}                   & \multicolumn{1}{c|}{503,780}                \\ \hline
\multicolumn{1}{|c|}{She/They}       & \multicolumn{1}{c|}{158,025}                 & \multicolumn{1}{c|}{508}                   & \multicolumn{1}{c|}{463,599}                \\ \hline
\multicolumn{1}{|c|}{They}           & \multicolumn{1}{c|}{132,374}                 & \multicolumn{1}{c|}{560}                   & \multicolumn{1}{c|}{506,064}                \\ \hline
\multicolumn{1}{|c|}{He/They}        & \multicolumn{1}{c|}{77,951}                  & \multicolumn{1}{c|}{514}                   & \multicolumn{1}{c|}{469,328}                \\ \hline
\multicolumn{1}{|c|}{She/He/They}    & \multicolumn{1}{c|}{20,882}                  & \multicolumn{1}{c|}{557}                   & \multicolumn{1}{c|}{611,227}                \\ \hline
\multicolumn{1}{|c|}{They/Xe}        & \multicolumn{1}{c|}{1,312}                   & \multicolumn{1}{c|}{468}                   & \multicolumn{1}{c|}{462,775}                \\ \hline
\multicolumn{1}{|c|}{He/They/Xe}     & \multicolumn{1}{c|}{1,015}                   & \multicolumn{1}{c|}{377}                   & \multicolumn{1}{c|}{387,722}                \\ \hline
\multicolumn{1}{|c|}{\textbf{Total}}     & \multicolumn{1}{c|}{\textbf{2,047,388}}                   & \multicolumn{1}{c|}{\textbf{4,051}}                   & \multicolumn{1}{c|}{\textbf{3,868,757}}
\\ \hline

\end{tabular}
\caption{Pronoun Group composition within collected sample. Users shows number of unique users in sample, tweets shows total tweets.}
\label{tab:sample}
\vspace{-3em}
\end{table}

\paragraph{Reply Collection}
We randomly sample 100 users from each pronoun group in our data set and collect up to 10 replies to their original tweets (not a reply). We collect the first 10 replies with the same \verb|conversation_id|. This results in 95,381 replies from 29,537 unique \verb|conversation_id|s. The reply sample is small due to complications in Twitter API access resulting from Twitter's change in leadership.

\paragraph{Toxicity Inference} 
To measure toxicity, we use \textit{Detoxify} model called \textit{unbiased}, a RoBERTa model fine-tuned on the Kaggle Jigsaw Unintended Bias in Toxicity Classification Challenge data set, as described above. This model has an AUC score of 92.11.

\begin{figure*}[h!]
\centering
\includegraphics[width=0.96\textwidth]{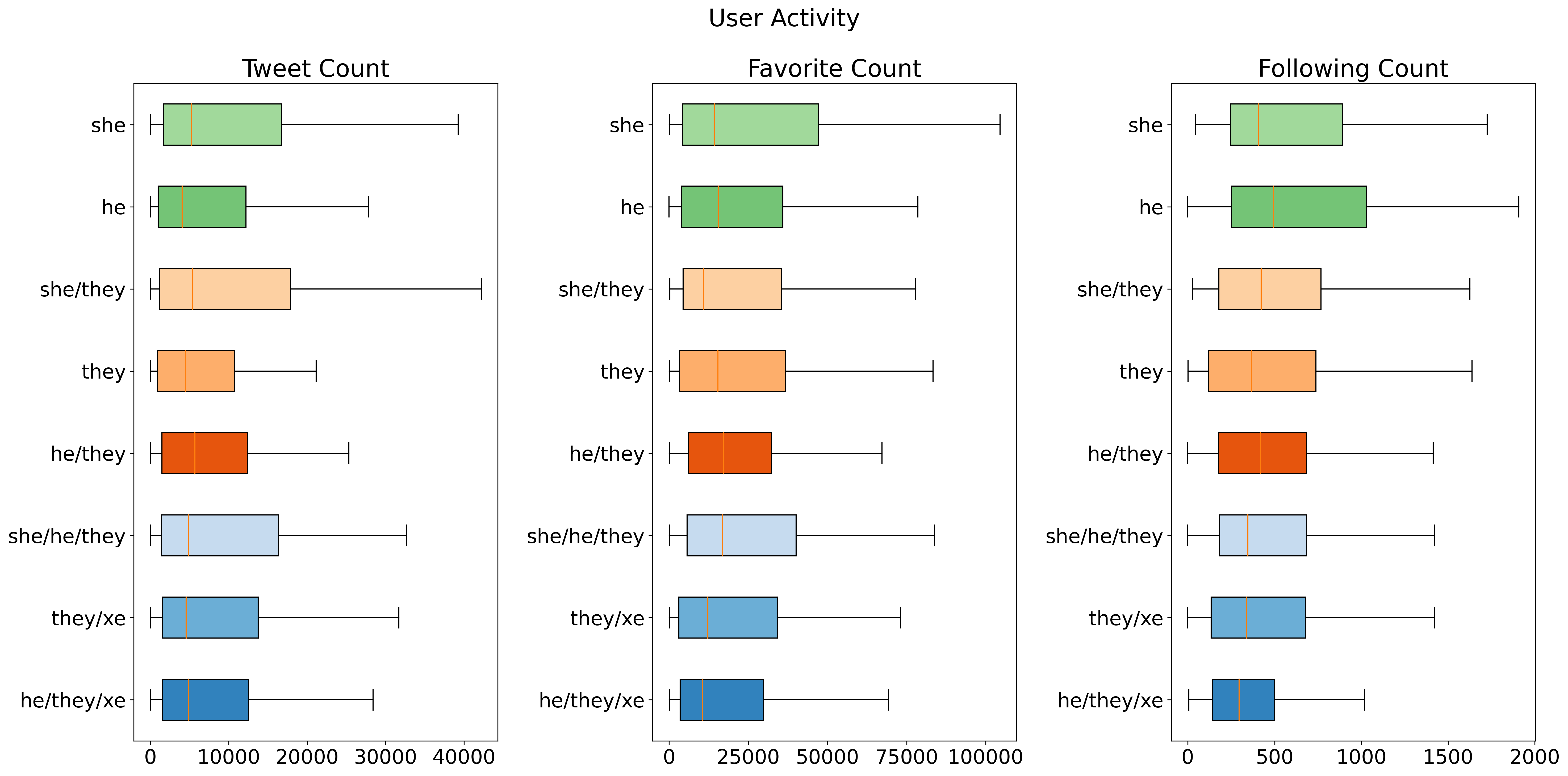}
\caption{User activity by pronoun group. Number of followers and favorites remains fairly similar between pronoun groups.}
\label{fig:activity}
\vspace{-7mm}
\end{figure*}
\section{Results}
\subsection{RQ1: Activity and Attention}
\textit{Activity} refers to a user’s engagement on Twitter, including the number of status updates (original tweets), favorites tweets (likes), and accounts followed. Overall, we find that user activity measured by likes, retweets and following count is relatively similar between binary and non-binary users, with only minor variations.
Figure~\ref{fig:activity} shows the activity distribution for each user pronoun group, with outliers excluded to highlight differences between groups. Tweet count denotes the number of original tweets, retweets and replies by a user. 
In Figure~\ref{fig:activity}, the pronoun group with the highest median tweet count is \textit{he/they} (5,676) and the lowest is \textit{they/xe} (4,553), There is a small difference in the number of tweets sent out between users with binary and non-binary pronouns overall (T-test statistic = 11.784, p $<$ 0.01) but the number is relatively small for both groups. 
The favorite count, denoting number of tweets liked by the user, has a slight difference in the number of tweets liked by accounts with non-binary versus binary pronouns in their profile (T-test statistic = 11.599, p $<$ .01). The pronoun groups with the highest median favorite count are \textit{he/they} (17,080), while lowest is \textit{they/xe} (12,214).
The following count, representing the number accounts a user follows, also has a very small difference between binary and non-binary users (T-statistic = 5.598, p $<$ .01). The pronoun group with the highest median following count is \textit{he} (494) and the lowest is \textit{she/he/they} (346).  Overall, user activity measured by likes, retweets and following count is relatively similar between binary and non-binary users, with only minor variations.

\textit{Attention} refers to the level of engagement and prominence users hold on Twitter, measured here by average number of retweets, likes, followers, and the percentage of verified users.
Based on these measures, we find that pronoun groups with less representation receive less attention than those with higher overall representation.


\begin{figure*}[h!]
\centering
\includegraphics[width=.96\textwidth]{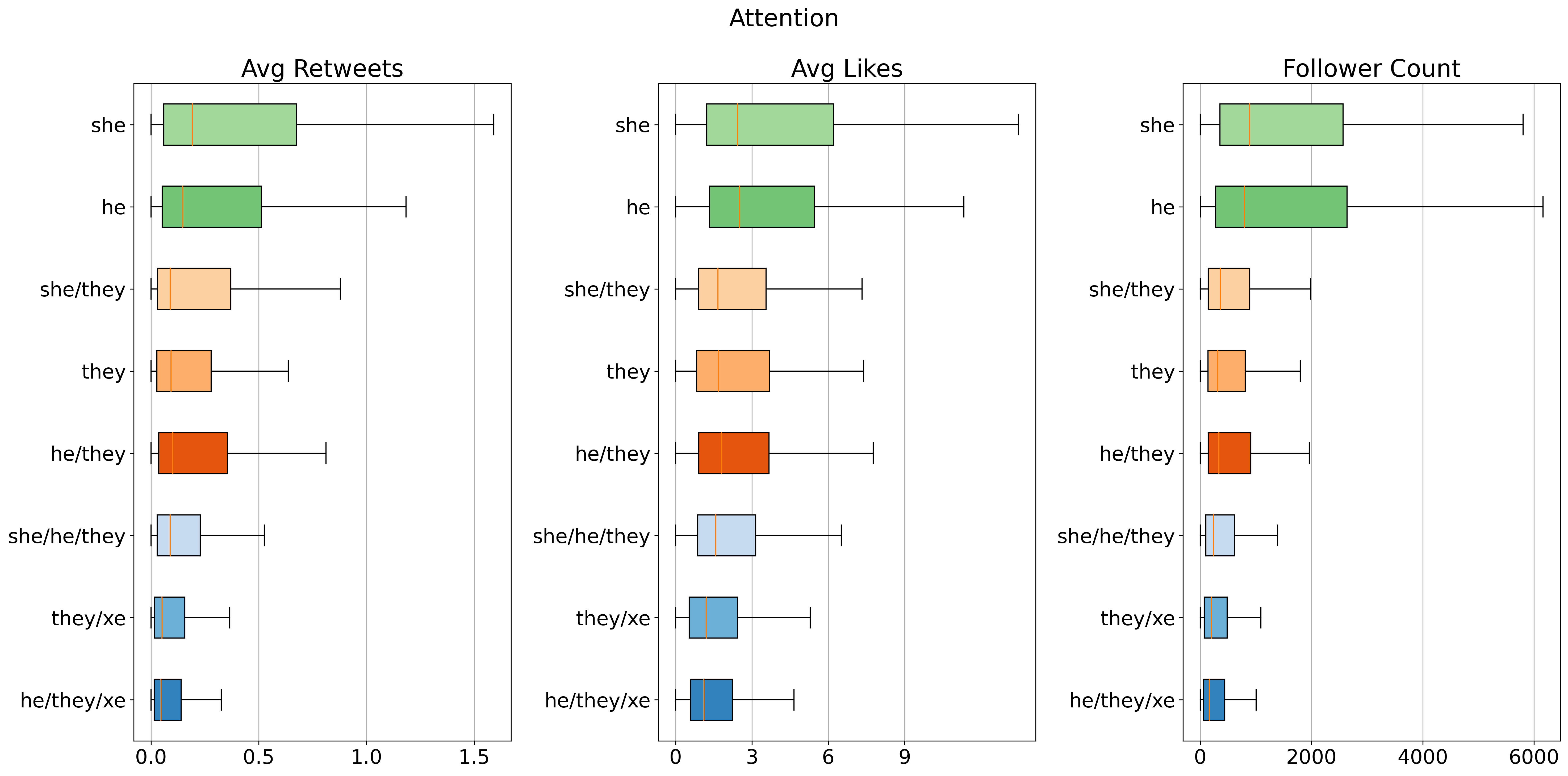}
\caption{Attention. Users with more representation receive more attention in retweet and like averages, as well as followers.}
\label{fig:attention}
\end{figure*}

In Figure~\ref{fig:attention} we observe pronoun group attention distributions. We observe the larger groups receive higher median retweets (\textit{she} (.192), \textit{he} (.148), and \textit{he/they} (.102)) while
smaller groups receiving lower median retweets (\textit{he/they/xe} (.045), \textit{they/xe} (.051) and \textit{she/he/they} (.089). We observe a similar pattern for median likes. The largest groups receive higher median likes (\textit{he} (2.510), \textit{she} (2.425) and \textit{he/they} (1.796) and \textit{he/they} (.102)) while the smaller groups receive lower median likes (\textit{he/they/xe} (1.111), \textit{they/xe} (1.207) and \textit{she/he/they} (1.584)). For both retweets and likes, there is a correlation between the representation of a pronoun group in the original dataset and the average number of likes received (Spearman = -.215, p $<$ 0.01, and Spearman = -.229, p $<$ 0.01, respectively). Number of followers shows an extreme version of this trend (Spearman = -.363, p $<$ 0.01). 

\subsection{RQ3: Toxicity in Tweets and Replies}


We next look at toxicity of tweets that a user posted, both original tweets and replies (sent tweets), and toxicity of the replies user received (received tweets).

\begin{figure}[h!]
\centering
\begin{subfigure}[]{}
\includegraphics[width=0.45\textwidth]{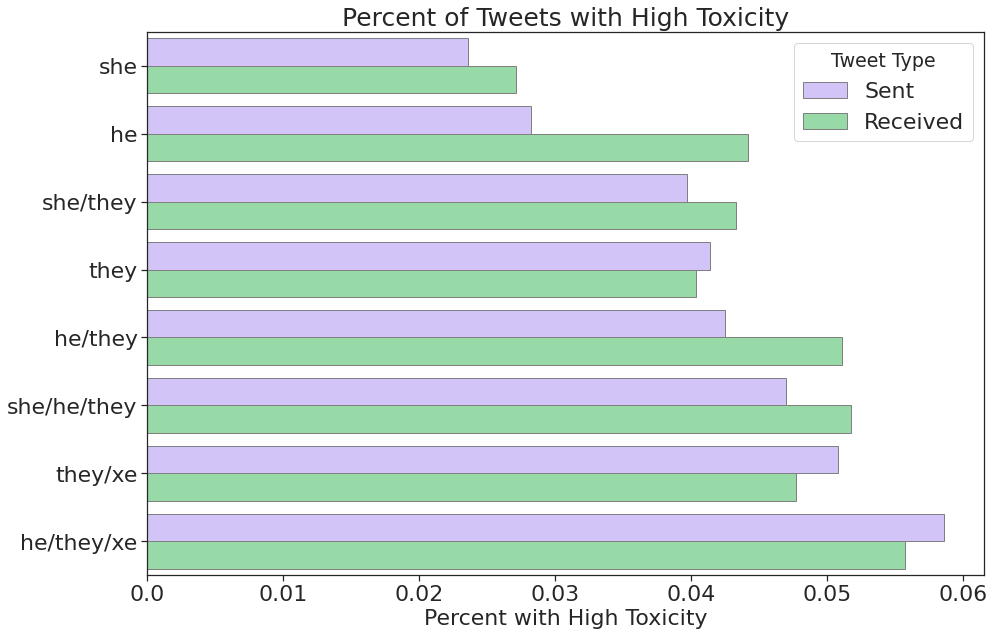}
   \label{fig:high_toxic} 
\end{subfigure}
\begin{subfigure}[]{}
\includegraphics[width=0.45\textwidth]{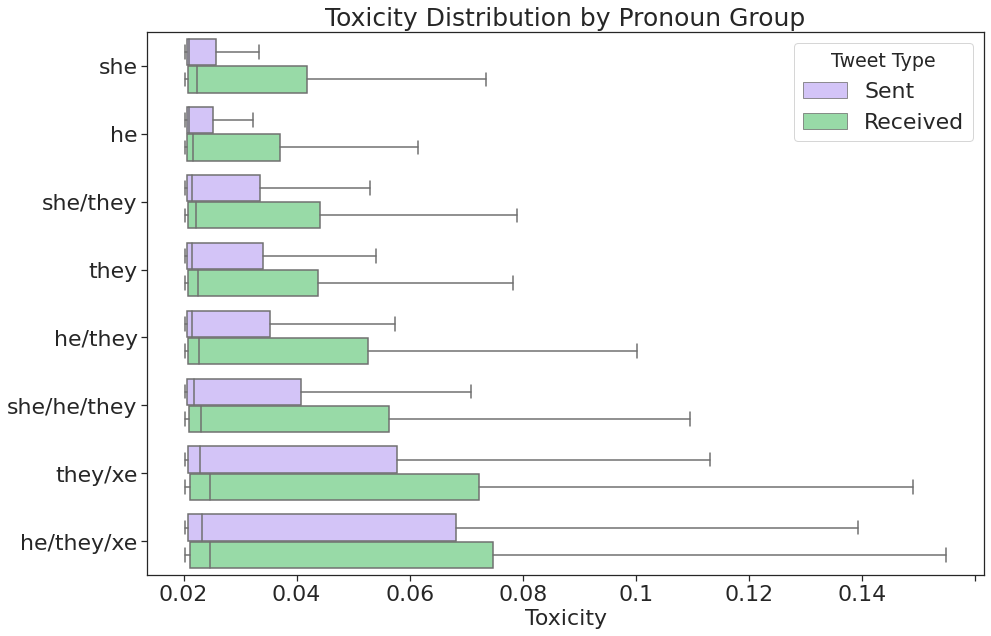}
   \label{fig:tox_distrib}
\end{subfigure}
\caption{(a) Percent of Tweets Posted and Percent of Replies Received Labeled as Toxic (toxicity $>$ 0.9). (b) Toxicity distribution for tweets posted and received. Tweets from non-binary have higher toxicity scores than those from binary users.}
\end{figure}

First, we look at the incidence of highly toxic tweets, with toxicity scores above some threshold. Content moderation algorithms may automatically flag such comments. 
Figure~\ref{fig:high_toxic}  plots the share of highly toxic tweets (toxicity $> 0.9$) among the sent and received tweets. 
All groups send out tweets where less than 1\% are highly toxic.
Five out of eight pronoun groups receive more highly toxic tweets than they send. The three groups which send out more highly toxic content than they receive are \textit{they/xe}, \textit{he/they/xe} and \textit{they}.
The two groups with the largest difference $|Toxicity(Received) - Toxicity(Sent)|$ are \textit{he} (.016) and \textit{he/they} (.009). Those with the least difference are \textit{they} (.001) and \textit{he/they/xe} (.003).
Surprisingly, the non-binary pronoun groups appear to post more highly toxic tweets than the binary groups. This suggests that more of their tweets would be flagged or removed by content moderation. These groups also appear to receive more highly toxic tweets.


In Figure \ref{fig:tox_distrib} we look at the distribution of toxicity scores for non-outliers. An outlier is defined as points outside of $[Q1 - 1.5*IQR, Q3 + 1.5*IQR]$, where Q$i$ is quarter $i$, and $IQR$ stands for inter-quartile range. 
In looking at the overall toxicity distributions, no pronoun group has a higher median toxicity sent score than median toxicity received score. 
The two groups \textit{he/they/xe} and \textit{they/xe} have the highest median toxicity score for tweets they send (.0032, .0028) and receive (.0048, .0047).
The groups \textit{he} and \textit{she} have the lowest median toxicity sent (.0009, .0010) and received (.0016, .0023). 
The groups with the highest difference in median $|Toxicity(Sent) - Toxicity(Received)|$ are \textit{they/xe} (.0019), \textit{he/they/xe} (.0015) and \textit{she/he/they} (.0014).
The groups with the lowest difference are \textit{he} (.0007) and \textit{he/they} (.0008).


There is a significant correlation between pronoun groups with large representation and the toxicity scores assigned to a tweet (Spearman = .16, p $<$ 0.01). Additionally, the toxicity scores for binary and non-binary users are significantly different (T-statistic = -125.72, p $<$ .01). In general, there is correlation between the group's share of highly toxic tweets received and highly toxic tweets sent.

\begin{table}[]
\small
\begin{tabular}{|c|p{0.7\linewidth}|c|}
\hline
\textbf{User} & \multicolumn{1}{c|}{\textbf{Tweet}}                                                        & \textbf{Toxicity} \\ \hline
binary        & RT @user: i’ll die for my niggas, i ride my niggas                                                    & .996              \\ \hline
binary        & 
Solid, got my first death threat today! Filled with such language as “faggot”,  “I will shoot you in themotherfucking mouth”, “you dumb ass…”
& .998              \\ \hline
non-binary    & I will start t one day One Fuckin Day                                                                                                                                                   & .981              \\ \hline
non-binary    & these fucking ladybug cockroach monsters will be the death of me                                                                                                                        & .998              \\ \hline
\end{tabular}
\caption{Examples of tweets flagged as highly toxic. Usernames are swapped with \textit{user} to preserve privacy.}
\label{tab:examples}
\vspace{-2em}
\end{table}

\begin{table}[h]
\centering

\footnotesize
\begin{tabular}{|c|c|c|c|}
\hline
\textbf{Group}       & \textbf{Profane} & \textbf{Profane \& Toxic} & \textbf{Toxic} \\ \hline
\textbf{she}         & 65260 (.171)     & 4797 (.013)               & 8986 (.024)    \\ \hline
\textbf{he}          & 71919 (.171)     & 6343 (.015)               & 11868 (.028)   \\ \hline
\textbf{she/they}    & 56490 (.149)     & 9150 (.024)               & 15045 (.039)   \\ \hline
\textbf{they}        & 65223 (.154)     & 10652 (.025)              & 17560 (.041)   \\ \hline
\textbf{he/they}     & 55788 (.154)     & 9480 (.026)               & 15385 (.042)   \\ \hline
\textbf{she/he/they} & 61935 (.146)     & 12223 (.029)              & 19982 (.047)   \\ \hline
\textbf{they/xe}     & 42997 (.139)     & 9268 (.029)               & 15751 (.051)   \\ \hline
\textbf{he/they/xe}  & 42455 (.142)     & 10531 (.035)              & 17589 (.059)   \\ \hline
\end{tabular}
\caption{Percent of tweets featuring profanities (Profane), with toxicity levels above 0.9 (Toxic) and those featuring both profanities and with toxicity levels above 0.9 (Profane \& Toxic).}
\label{tab:prof}
\vspace{-3em}
\end{table}

In an effort to understand the difference in toxicity detection results for binary and non-binary users we sifted through some tweets flagged as highly toxic. We observed a pattern in which non-toxic tweets with profanities (curse words) present obtain a high toxicity score. We list some examples of this in Table \ref{tab:examples}.
We then performed a preliminary exploration of the overlap between toxicity and profanity using a data set of profane language\footnote{https://github.com/surge-ai/profanity.git}.

In Table \ref{tab:prof} we report our findings regarding the overlap between profanity and toxicity.
In the left-most column (Profane) we report each pronoun group's overall presence of profanities in their tweets. Numbers are represent as $N (P)$, where $N$ denotes the number of profanities present and $P$ denotes the percentage of overall tweets featuring profanities. We observe that tweets from users with binary pronouns have the highest rates of profanities.
In the right-most column (Toxic) we report the amount of tweets labeled as highly toxic (toxicity $>$ .9). As discussed earlier, the groups with the lowest initial representation have the highest toxicity scores.
In the middle column (Profane \& Toxic) we look at amounts of tweets that are labeled as highly toxic and contain at least one profanity.
We see that the groups with the highest use of profanities have the lowest scores for Profane \& Toxic. 
This suggests no link between the presence of profanities and the toxicity score assigned.

In conclusion, posts from non-binary users are flagged as toxic at higher rates than binary users. This does not seem to depend on the use of profanities. We would be interested in further analysis into the link between high toxicity scores and pronoun groups.
These results show a surprising divergence between tweets with profanity and those classified as toxic.

\section{Conclusion}
In this study, we conduct an exploratory analysis of online behavior of eight pronoun groups, including both binary and non-binary pronoun series. We investigate group behavior online by analyzing activity, attention, emotions, and toxicity. We also examine the levels of toxicity and types of emotions expressed in replies to  tweets by the pronoun groups.

We find that non-binary pronoun groups with less overall representation on Twitter receive less attention via retweets, likes and followers than groups with higher overall representation. These low-response users engage on Twitter's platform by sending and favoriting tweets at about an equal rate as other users.

We find that, in comparison with Twitter users with binary pronouns, non- binary groups have higher levels of toxicity detected in their tweets. 
It is surprising because users with non-binary pronouns use less profanity, even though their tweets are classified as more toxic. We hypothesize that the toxicity classifier may contain  \textit{dialect bias} where dialect used in queer communities is falsely flagged as expressing toxicity. This fits into prior work showing how social media content by drag queens are overly classified as hate speech \cite{fightinghatespeech}.



There are several limitations to this work. 
The analysis is limited by the specificity of the population at hand, excluding non-binary Twitter users without pronouns in biographies. Pronoun order is unaccounted for: \textit{she/they} is treated the same as \textit{they/she}.
We select users for the study from a single COVID-19 data set, though could ensure better representation of active Twitter users by beginning with multiple data sources.
Additionally, the number of replies collected per tweet is fairly low (n = 10). 
We are interested in seeing how results change if we collect more replies.


Another promising avenue of future research could analyze how toxicity detectors handle text written by individuals identifying as non-binary. Such detectors’ results could be compared to those based on crowd-sourced annotators and true labels manually determined by experts. In addition, researchers could examine the network characteristics associated with different user pronouns. These are just a few examples of the possibilities raised by research into the relationship between gender expression, as shown through pronouns, and behaviors observed on social media.


\subsubsection*{Acknowledgements}
We are grateful to Siyi Guo and Donald Berry for helping with initial analysis.

%
%
%
%
\bibliographystyle{splncs04}
\bibliography{sample-base}

\end{document}